\documentclass[aps,pra,twocolumn,superscriptaddress,floatfix,footinbib,notitlepage,groupaddress,showpacs]{revtex4-1} 
\usepackage{times,color,amsfonts,amssymb,amsbsy,amsthm,amsmath,graphics,graphicx,hyperref,bm,bbm,makecell,mathtools,dsfont}
\usepackage{upgreek,hyperref}
\newcommand{\ignore}[1]{}

\hypersetup{colorlinks,linkcolor={blue},citecolor={blue},urlcolor={blue}}
\newcommand{\ket}[1]{\left|{#1}\right\rangle}
\newcommand{\bra}[1]{\left\langle{#1}\right|}

\newcommand{\ketbra}[2]{|{#1}\rangle\langle{#2}|}
\newcommand{\MGH}[1]{\textcolor{blue}{#1}}
\newcommand{\ignor}[1]{}

\DeclareFontFamily{OT1}{pzc}{}
\DeclareFontShape{OT1}{pzc}{m}{it}%
              {<-> s * [1.25] pzcmi7t}{}
\DeclareMathAlphabet{\mathpzc}{OT1}{pzc}%
                                 {m}{it}
\let\oldsqrt\sqrt
\def\sqrt{\mathpalette\DHLhksqrt}
\def\DHLhksqrt#1#2{%
\setbox0=\hbox{$#1\oldsqrt{#2\,}$}\dimen0=\ht0
\advance\dimen0-0.2\ht0
\setbox2=\hbox{\vrule height\ht0 depth -\dimen0}%
{\box0\lower0.4pt\box2}}
\begin{document}

\title{Scheme for coherent-state quantum process tomography via normally-ordered moments}

\author{M. Ghalaii}
\email{ghalaiim@gmail.com}
\affiliation{School of Electronic and Electrical Engineering, University of Leeds, Leeds LS2 9JT, United Kingdom}
\affiliation{Department of Physics, Sharif University of Technology, Tehran 14588, Iran}

% [A] removed: SRK

\author{A. T. Rezakhani}
\affiliation{Department of Physics, Sharif University of Technology, Tehran 14588, Iran}

\begin{abstract}
Using coherent states in optical quantum process tomography is a practically-relevant approach. Here, we develop a framework for complete characterization of quantum-optical processes in terms of normally-ordered moments by using coherent states as probes. We derive the associated superoperator tensors for several optical processes. We also show that our technique can be used to determine nonclassicality features of quantum-optical states and processes. Furthermore, we investigate identification of multi-mode Gaussian processes and show that the number of necessary probe coherent states scales linearly with the number of modes. 
\end{abstract}

\pacs{42.30.Wb, 03.65.Wj, 42.50.Gy}
% [A] double-check the pacs to ensure their relevance.
\date{\today}

\maketitle

%%%%%%%%%%%%%%%%%%%%%%%%%%%%%%%%%%%%%
\section{Introduction} 
\label{sec:int}

Quantum process tomography (QPT) is used to characterize an unknown quantum process/operation \cite{Nielsen,Chuang,Poyatos}. Several methods have been proposed for QPT, e.g., standard QPT \cite{Nielsen,Chuang,Poyatos}, ancilla-assisted process tomography \cite{Leung,D'Ariano-AAPT-1,Altepeter,D'Ariano-AAPT-2}, and direct characterization of quantum dynamics \cite{Mohseni-DCQD} (for a review see Ref.~\cite{Mohseni},). 

Quantum-optical operations are of special interest because, for example, quantum optics provides a promising platform for performing various physical tasks, testing quantum computation, and performing secure quantum communications \cite{Knill,Gisin,O'Brien,Childs,Mitchell}. However, realization of such systems may be difficult due to necessity of generating probe states that are highly nonclassical. Some recently proposed QPT methods alleviate parts of this obstacle by using coherent states as probes \cite{Lobino,Rahimi-Keshari-qpt,Blandino,Anis,Kumar,Wang-csQPT}. In other words, a process can be experimentally characterized by analyzing its effect on a set of coherent states. Coherent-state quantum process tomography (csQPT) offers favorable features such as experimental feasibility within current technology---e.g., optical homodyne detection allows straightforward identification of output states \cite{Leonhardt,Lvovsky}.

In quantum-optical computation and communication, it is important to analyze various features of output states out of an unknown quantum process, such as their nonclassicality, or to determine how specific nonclassical effects, such as squeezing or antibunching \cite{Vogel-Book}, are generated or affected by quantum processes. Despite much progress \cite{Rahimi-Keshari-qpt,Lobino,Rahimi-Keshari}, however, these tasks are still formidable.

It has been known that by comparing \textit{normally-order moments} of input and output states allows one to obtain nonclassical features of a quantum-optical process \cite{Shchukin-1,Shchukin-2}. Moreover, there are quantum states, such as Gaussian states, which can be uniquely represented by a finite number of moments \cite{Ferraro,Adesso} (despite that their density matrices in the Fock basis are infinite dimensional). Thus, for Gaussian processes \cite{Weedbrook}, it seems more reliable to perform csQPT in terms of normally-ordered moments rather than the Fock basis. Interestingly as well, normally-ordered moments can be experimentally characterized by using homodyne correlation measurements comprised of beam splitters and photon detectors \cite{Shchukin-2}. And in the case of multi-mode Gaussian processes, some recent experimental measurement techniques have also been proposed \cite{Pinel,Opatrny} .

In this paper, we propose a method for csQPT which is based on normally-ordered moments. We demonstrate analytically a scheme by which an unknown quantum process can be completely identified based on measuring output moments. The paper is structured as follows. In Sec.~\ref{Formalism}, we explain the formalism and work out the relation between output and input moments. A general relation is obtained for rank-$4$ superoperator elements for both single- and multi-mode optical processes. In Sec.~\ref{Examples}, the method is illustrated by studying some processes of interest in quantum optics and quantum information. The evolution of nonclassicality is also discussed through several examples. In particular in Sec.~\ref{gaussian Processes}, we characterize multi-mode Gaussian processes. The paper is summarized in Sec.~\ref{Conclusion}.

%%%%%%%%%%%%%%%%%%%%%%%%%%%%%%%%%%%%%
\section{Formalism: Superoperator in the normally-ordered moment basis}
\label{Formalism}

\begin{figure}[tp]
\includegraphics[scale=0.35]{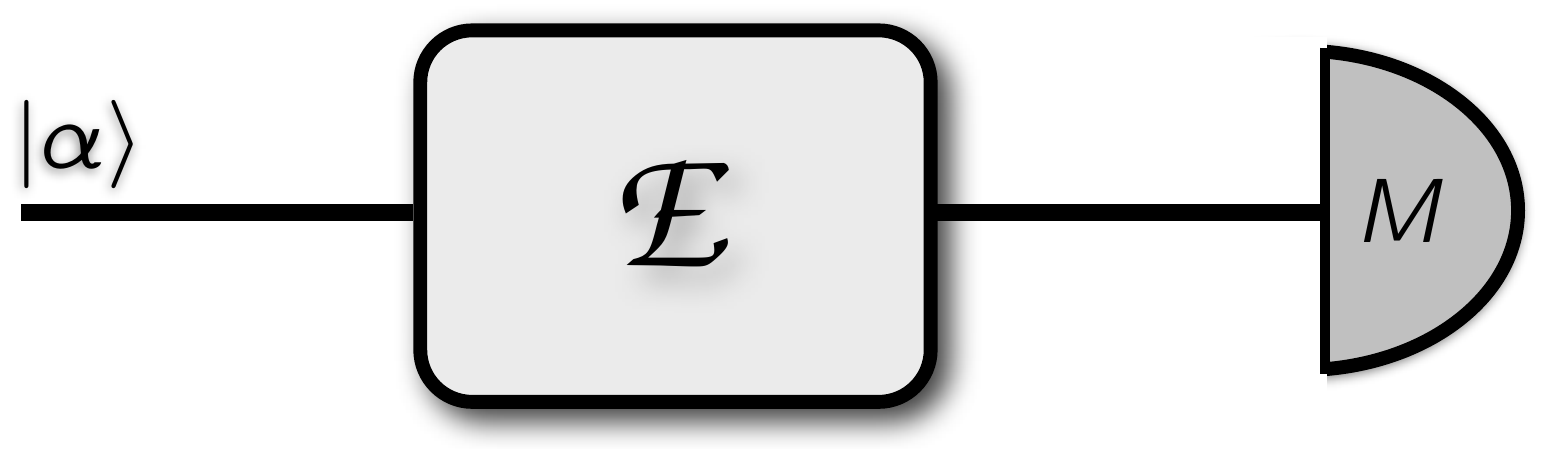}
\caption{Schematic presentation of coherent-state quantum process tomography. Here $M$ denotes moment measurements.}
\label{csqpt}
\end{figure}

In this section, we show that by applying a quantum process $\mathpzc{E}$ to a set of input coherent states and measuring the output states, one can characterize the process---Fig.~\ref{csqpt}. Any quantum state $\widehat{\varrho}$ can be expressed as a mixture of coherent states in the following form:
\begin{equation}
\label{Glub-Sud-rep}
\widehat{\varrho}=\int_{\mathds{C}} \mathrm{d}^2\upalpha \,P_{\widehat{\varrho}}(\upalpha) \ketbra{\upalpha}{\upalpha},
\end{equation} 
where $P_{\widehat{\varrho}}(\upalpha)$ is the Glauber-Sudarshan $P$ function of the state, $\ket{\upalpha}$ is a coherent state, and the integration is over the whole complex plane \cite{Glauber,Sudarshan}. By using Eq.~\eqref{Glub-Sud-rep}, we find a relation between the $P$ functions of the input state $\widehat{\varrho}$ and the output state $\mathpzc{E}[\widehat{\varrho}]$,
\begin{equation}
\label{P-func-proc}
P_{\mathpzc{E}[\widehat{\varrho}]}(\upbeta) = \int_{\mathds{C}} \mathrm{d}^2 \upalpha\, P_{\widehat{\varrho}}(\upalpha) P_{\mathpzc{E}[\ketbra{\upalpha}{\upalpha}]}(\upbeta |\upalpha),
\end{equation}
where $P_{\mathpzc{E}[\ketbra{\upalpha}{\upalpha}]}(\upbeta |\upalpha)$ is the $P$ function of the output state for an input coherent state $\ket{\upalpha}$, and $P_{\widehat{\varrho}}(\upalpha)$ and $P_{\mathpzc{E}[\widehat{\varrho}]}(\upbeta)$ are, respectively, $P$ functions of the input and output states. We remind the definition of normally-ordered moments, 
\begin{equation}
M_{jk}(\widehat{\varrho})=\mathrm{Tr}[\widehat{\varrho}~\widehat{\mathrm{a}}^{\dag k} \widehat{\mathrm{a}}^j],
\label{1-moments}
\end{equation} 
for the quantum state $\widehat{\varrho}$ \cite{Cahill-Glauber,Glauber}. Multiplying both sides of Eq.~\eqref{P-func-proc} by $\upbeta^{\ast k} \upbeta^j$ and integrating over the entire complex plane yields
\begin{equation}
\label{m-m}
M_{jk}(\mathpzc{E}[\widehat{\varrho}])= \int_{\mathds{C}} \mathrm{d}^2 \upalpha\, P_{\mathrm{in}}(\upalpha)M_{jk}(\mathpzc{E}[\ketbra{\upalpha}{\upalpha}]),
\end{equation}
where 
\begin{align}
M_{jk}(\mathpzc{E}[\widehat{\varrho}])&=\mathrm{Tr}\left[\mathpzc{E}[\widehat{\varrho}] ~{\widehat{\mathrm{a}}^{\dagger k}} \widehat{\mathrm{a}}^j \right],\label{mom-out-coh}\\
M_{jk}(\mathpzc{E}[\ketbra{\upalpha}{\upalpha}])&=\mathrm{Tr}\left[\mathpzc{E}[\ketbra{\upalpha}{\upalpha}]~ {\widehat{\mathrm{a}}^{\dagger k}} \widehat{\mathrm{a}}^j \right]
\end{align}
are the normally-ordered moments of the output state for the input states $\widehat{\varrho}$ and the input coherent state $\ket{\upalpha}$, respectively. 
% [A] P_{in} has not been defined here.

%---------------------------------------------------------------------------------------
%---------------------------------------------------------------------------------------

Next substituting the power series expansion 
% [A]
%of $M_{jk}(\mathpzc{E}[\ketbra{\upalpha}{\upalpha}])$,
\begin{align}
M_{jk}(\mathpzc{E}[|\upalpha\rangle\langle\upalpha|])=&\sum_{mn}\partial_{\upalpha}^m
 \partial_{\upalpha^\ast}^{n}M_{jk}(\mathpzc{E}[\ketbra{\upalpha}{\upalpha}])\Big|_{\upalpha=0} \nonumber \\
 &\times \frac{\upalpha^m \upalpha^{\ast n}} {m! n!}
\end{align}
(which is convergent for all $\upalpha$s---Appendix~\ref{conv-of-M}) in Eq.~\eqref{m-m} gives the following relation between output and input normally-ordered moments:
\begin{equation}
M_{jk}(\mathpzc{E}[\widehat{\varrho}])=\sum_{mn}\mathpzc{E}^{mn}_{jk}M_{mn}(\widehat{\varrho}),
\label{mom-in-out}
\end{equation}
where the rank-$4$ \textit{superoperator} tensor $\mathpzc{E}^{mn}_{jk}$ is given by
\begin{equation}
\label{sup}
\mathpzc{E}^{mn}_{jk}:=\frac{1}{m!n!}\partial_{\upalpha}^m
 \partial_{\upalpha^\ast}^{n}M_{jk}(\mathpzc{E}[\ketbra{\upalpha}{\upalpha}])\Big|_{\upalpha=0}.
\end{equation}
Thus, in principle, by using different coherent states $\ket{\upalpha}$ as probes of the process and measuring normally-ordered moments of the output states, $M_{jk}(\mathpzc{E}[\ketbra{\upalpha}{\upalpha}])$, one can evaluate the process tensor elements $\mathpzc{E}^{mn}_{jk}$ by taking partial derivatives of the output moments $M_{jk}(\mathpzc{E}[\ketbra{\upalpha}{\upalpha}])$ with respect to $\upalpha$ and $\upalpha^\ast$, which are estimated from experimental measurement and computed at $\upalpha=0$.
% [M] modified:
This moment measurement can be achieved by the experimental setups proposed in Refs. \cite{Shchukin-2,Opatrny,Pinel}.

Additionally, we can find the tensor elements $\mathpzc{E}^{mn}_{jk}$ in terms of the elements of the representation of the quantum process in the Fock basis ($\mathcal{E}_{jk}^{mn}$) as
\begin{equation}
\mathpzc{E}_{jk}^{mn}=\sum_{\ell,s=0}^{\infty}\frac{(-1)^s}{\ell!}\sqrt{\frac{(\ell+j)!(\ell+k)!}{(m-s)!(n-s)!}} \mathcal{E}_{j+\ell,k+\ell}^{m-s,n-s},
\end{equation}
and conversely---see Appendix~\ref{conv-of-M}.

% [A] removed; seems good information, but not necessary
\ignore{
We note that for the input coherent state $\ket{\upalpha}$, Eq.~\eqref{mom-in-out} can also be written in the matrix form
\begin{equation}
\label{mom-mat}
M(\mathpzc{E}[\ketbra{\upalpha}{\upalpha}]) = \mathpzc{E} M(\ketbra{\upalpha}{\upalpha}),
\end{equation}
from whence one could have argued---provided that one has the moment matrices of the input and output states---that one could find $\mathpzc{E}$ by inverting Eq.~\eqref{mom-mat},
\begin{equation}
\mathpzc{E}=M(\mathpzc{E}[\ketbra{\upalpha}{\upalpha}])  M^{-1}(\ketbra{\upalpha}{\upalpha}),
\end{equation}
where the matrix $M(\ketbra{\upalpha}{\upalpha})$ is generated by sufficient different input coherent states. This, however, is impossible because for the case of ``coherent-state'' QPT the inverse matrix $M^{-1}(\widehat{\varrho})$ does not exist---$M(\ketbra{\upalpha}{\upalpha})$ is a rank-$1$ matrix, hence possessing zero eigenvalues. 
}

Alternatively, by employing the normally-ordered characteristic function of the state $\widehat{\varrho}$,
\begin{equation}
\Lambda_{\widehat{\varrho}}(\upxi)=\mathrm{Tr}[\widehat{\varrho} ~\mathrm{e}^{\upxi \widehat{\mathrm{a}}^{\dag}}  \mathrm{e}^{-\upxi^\ast\widehat{\mathrm{a}}}],
\end{equation}
Eq.~\eqref{sup} can be written as
\begin{equation}
\label{sup-alternate}
\mathpzc{E}^{mn}_{jk}=\frac{(-1)^j}{m!n!}\partial_{\upalpha}^m
\partial_{\upalpha^\ast}^{n} \partial_{\upxi}^k  \partial_{\upxi^\ast}^{j} \Lambda_{\mathpzc{E}[\ketbra{\upalpha}{\upalpha}]}(\upxi|\upalpha)\Big|_{\upalpha,\upxi=0},
\end{equation}
where $\Lambda_{\mathpzc{E}[\ketbra{\upalpha}{\upalpha}]}(\upxi|\upalpha)$ is the characteristic function of the $P$ function of the output state for the input coherent state $\ket{\upalpha}$. Hence, when the characteristic function is available, one can calculate the tensor elements by using Eq.~\eqref{sup-alternate}. As an example, later in Sec.~\ref{decoherence}, we identify the process of decoherence for a squeezed vacuum state by using its time-variant characteristic function. 

Multi-mode generalization of the above characterization method is straightforward. In the $\mathpzc{m}$-mode case, consider $\mathbf{n}=(n_1,\ldots,n_{\mathpzc{m}})$, $\mathbf{m}=(m_1,\ldots,m_{\mathpzc{m}})$, and assume the multi-mode coherent states $|\bm{\upalpha}\rangle=|\upalpha_1,\ldots,\upalpha_{\mathpzc{m}}\rangle$ as the probes. The superoperator is then given by
\begin{align}
\label{mult-sup}
\mathpzc{E}^{\mathbf{m}\mathbf{n}}
_{\mathbf{j}\mathbf{k}}=\prod_{s=1}^{\mathpzc{m}} \frac{1}{m_{s}!n_{s}!} \partial_{\upalpha_{s}}^{m_{s}} \partial_{\upalpha_{s}^{\ast}}^{n_{s}}
 M_{\mathbf{j}\mathbf{k}}(\mathpzc{E}[|\bm{\upalpha}\rangle \langle \bm{\upalpha}|]) \bigg|_{\upalpha_s=0},
\end{align}
where $ M_{\mathbf{j}\mathbf{k}}(\mathpzc{E}[|\bm{\upalpha}\rangle \langle \bm{\upalpha}|])$, assuming $\mathbf{j}=(j_1,\ldots,j_{\mathpzc{m}})$ and $\mathbf{k}=(k_1,\ldots,k_{\mathpzc{m}})$, is a ``super matrix," which includes all output moments of different modes. As an example of this type, later in Sec.~\ref{Examples-BS}, we derive the exact and closed-form expression of the superoperator tensor of a beam splitter process.

% [A] removed; redundant
\ignore{
To summarize, according to Eqs.~\eqref{mom-in-out} and \eqref{mult-sup}, by subjecting a set of input coherent states to an unknown process and measuring the normally-ordered moments of the output states, the superoperator $\mathpzc{E}^{mn}_{jk}$ can be estimated. 
}

% [M] text added (start)
Before we end this section, two brief remarks regarding measurements and comparing some methods are in order. (i) In the Fock-basis process tomography \cite{Rahimi-Keshari-qpt}, the coefficients of the output states (for input coherent state) should be measured in the Fock basis. Such measurements are usually based on homodyne detection (see Ref.~\cite{Lvovsky} for an extensive review). As the number of Fock states is in principle infinite, in practice one needs to truncate the associated Hilbert space to a cutoff Fock number. In this method two detectors are used to perform the homodyne measurement. (ii) The moment measurement technique proposed in Ref.~\cite{Shchukin-2} is a step-by-step method; in order to measure higher order moments one needs to have the information of the previous (lower) moments. For example, estimating $M_{21}$ requires using the information by which we evaluate lower ordered moments such as $M_{10}$ and $M_{11}$. In contrast with the Fock-basis tomography, here one requires more detectors to do the homodyne correlation measurements. For example, to measure moments $M_{jk}$ up to $j,k=n$, the number of detectors needed at the last step is $n$. Hence, in practice similarly to the Fock-basis method, imposing a cutoff will be unavoidable. However, if we are given that the process is Gaussian (or more generally, the number of moments of the output states is finite), in principle no cutoff would be required (and hence no error is incurred); whereas even for these special cases in the Fock-basis method one still needs to do the complete tomography of the output state (and hence should inevitably accept some extent of error).
% [M] text added (end)

%%%%%%%%%%%%%%%%%%%%%%%%%%%%%%%%%%%%%
\section{Examples}
\label{Examples}

In this section, based on Eqs.~\eqref{sup} and \eqref{mult-sup}, we demonstrate our tomography method by applying it to several important quantum-optical processes, for which we analytically derive the corresponding superoperator tensor elements $\mathpzc{E}_{jk}^{mn}$. See Table~\ref{table-1} for a summary of the results.

%%%%%%%%%%%%%%%%%%%%%%%%%%%%%%%%%%%%%
\subsection{Identity}

For the identity process, $\mathpzc{E}_{\mathrm{id}}[\ketbra{\upalpha}{\upalpha}]=\ketbra{\upalpha}{\upalpha}$, the normally-ordered moments of the output states are 
\begin{equation}
M_{jk}(\mathpzc{E}_{\mathrm{id}}[\ketbra{\upalpha}{\upalpha}])={\upalpha}^j\upalpha^{\ast k}.
\end{equation}
By inserting this elements into Eq.~\eqref{sup}, we find  $\mathpzc{E}^{mn}_{jk}=\updelta_{mj}\updelta_{nk}$, as one would expect.

%%%%%%%%%%%%%%%%%%%%%%%%%%%%%%%%%%%%%
\subsection{Attenuation (lossy channel)}

The effect of the process of attenuation on the coherent state $\ket{\upalpha}$ is given by $\mathpzc{E}_{\mathrm{att}}[\ketbra{\upalpha}{\upalpha}]=\ketbra{\eta\upalpha}{\eta\upalpha}$, where $0 < \eta <1$. The normally-ordered moments of the output states are
\begin{equation}
M_{jk}(\mathpzc{E}_{\mathrm{att}}[\ketbra{\upalpha}{\upalpha}])=\eta^{j+k}\upalpha^j\upalpha^{\ast k},
\end{equation}
whence $\mathpzc{E}^{mn}_{jk}=\eta^{j+k}\updelta_{mj}\updelta_{nk}$.

%%%%%%%%%%%%%%%%%%%%%%%%%%%%%%%%%%%%%
\subsection{Displacement}

The displacement operator is defined as $\widehat{D}(\widehat{\mathrm{a}}, \upbeta) =\mathrm{e}^{\upbeta \widehat{\mathrm{a}}^{\dagger}-\upbeta^\ast\widehat{\mathrm{a}}}$ \cite{Scully} 
%\begin{align}
%\widehat{D}(\widehat{\mathrm{a}}, \upbeta) &=\mathrm{e}^{\upbeta \widehat{\mathrm{a}}^{\dagger}-\upbeta^\ast\widehat{\mathrm{a}}}\nonumber\\
%&= \mathrm{e}^{-|\upbeta|^2/2}\mathrm{e}^{\upbeta \widehat{\mathrm{a}}^{\dagger}}\mathrm{e}^{-\upbeta^\ast\widehat{\mathrm{a}}},
%\end{align}
and its action on a coherent state is given by
% [M] the missing phase factor is added to the RHD of the Eq.  
\begin{equation}
\widehat{D}(\widehat{\mathrm{a}}, \upbeta)\ket{\upalpha}=e^{\frac{1}{2}(\upbeta \upalpha^{\ast} - \upbeta^{\ast}\upalpha )} \ket{\upalpha+\upbeta}.
\end{equation}
Thus, from Eqs.~\eqref{mom-out-coh} and \eqref{sup} we have 
\begin{equation}
M_{jk}(\mathpzc{E}_{\mathrm{disp}}[\ketbra{\upalpha}{\upalpha}])=(\upalpha+\upbeta)^j(\upalpha^\ast+\upbeta^\ast)^k.
\end{equation}
This yields
\begin{equation}
\label{disp-proc}
\mathpzc{E}_{jk}^{mn}=\frac{j!k!}{m!n!(j-m)!(k-n)!}\upbeta^{j-m}\upbeta^{\ast k-n}.
\end{equation} 
As expected, the tensor elements depend on the amount of displacement $\upbeta$.

%%%%%%%%%%%%%%%%%%%%%%%%%%%%%%%%%%%%%
\subsection{Photon subtraction and addition}

The two processes that enable us to remove or add a single photon in a beam light, are photon subtraction $\mathpzc{E}_{\mathrm{sub}}$ and photon addition $\mathpzc{E}_{\mathrm{add}}$ processes, respectively. The effect of these processes on input coherent states $\ket{\upalpha}$ are given by $\mathpzc{E}_{\mathrm{sub}}[\ketbra{\upalpha}{\upalpha}] =\widehat{\mathrm{a}}\ketbra{\upalpha}{\upalpha}\widehat{\mathrm{a}}^{\dagger}$ and $\mathpzc{E}_{\mathrm{add}}[\ketbra{\upalpha}{\upalpha}] =\widehat{\mathrm{a}}^{\dagger}\ketbra{\upalpha}{\upalpha}\widehat{\mathrm{a}}$, where $\widehat{\mathrm{a}}$ and $\widehat{\mathrm{a}}^{\dagger}$ are the field annihilation and creation operators, respectively. From Eq.~\eqref{mom-out-coh}, the normally-ordered moments of the output states for an input coherent state $\ket{\upalpha}$ are obtained as
\begin{align}
M_{jk}(\mathpzc{E}_{\mathrm{sub}}[\ketbra{\upalpha}{\upalpha}])=&\upalpha^{j+1}\upalpha^{\ast k+1},\\
M_{jk}(\mathpzc{E}_{\mathrm{add}}[\ketbra{\upalpha}{\upalpha}])=& kj\upalpha^{j-1}\upalpha^{\ast k-1}+\upalpha^{j+1}\upalpha^{\ast k+1}   \nonumber \\
&+(k+j+1)\upalpha^{j}\upalpha^{\ast k}.
\end{align}
The tensor elements obtained via Eq.~\eqref{sup} are
\begin{equation}
\label{tens-sub}
 \mathpzc{E}^{mn}_{jk}=\updelta_{m,j+1}\updelta_{n,k+1},
\end{equation}
for the photon subtraction process, and 
\begin{align}
 \mathpzc{E}^{mn}_{jk}=& kj\updelta_{m,j-1}\updelta_{n,k-1}+\updelta_{m,j+1} \updelta_{n,k+1}     \nonumber \\
 &+(k+j+1)\updelta_{mj}\updelta_{nk},
\end{align}
for the photon addition process.

\subsection{Beam splitter}
\label{Examples-BS}

Beam splitter is a two-mode optical process whose action on two input coherent states $\ket{\upalpha_1}$ and $\ket{\upalpha_2}$ is as fallows \cite{Rahimi-Keshari-qpt}:
\begin{align}
\mathpzc{E}_{\mathrm{BS}} [\ketbra{\upalpha_1, \upalpha_2}{\upalpha_1, \upalpha_2}] =&\ket{T\upalpha_1-R\upalpha_2, R\upalpha_1+T\upalpha_2} \nonumber \\
&\bra{T\upalpha_1-R\upalpha_2, R\upalpha_1+T\upalpha_2},
\end{align}
where $T$ and $R$ are, respectively, the transmissivity and reflectivity of the beam splitter. By using Eq.~\eqref{mult-sup} the superoperator tensor elements are obtained as
\begin{align}
\mathpzc{E}_{j_1j_2k_1k_2}^{m_1m_2n_1n_2} =\sum_{pr}\binom{j_1}{p}\binom{k_1}{r}\binom{j_2}{m_1-p}\binom{k_2}{n_1-r}  \nonumber \\
\times   T^{2p+2r+j_2+k_2-m_1-n_1}  R^{m_1+n_1+j_1+k_1-2p-2r} \nonumber \\
\times  (-1)^{j_1+k_1-p-r} \updelta_{m_1+m_2,j_1+j_2} \updelta_{n_1+n_2,k_1+k_2}, 
\end{align}
as an explicit function of $T$ and $R$, see Appendix~\ref{app-BS} for details.

%%%%%%%%%%%%%%%%%%%%%%%%%%%%%%%%
\subsection{Schr\"{o}dinger cat-state generation}

A Schr\"{o}dinger cat-state, $(|\upalpha \rangle +i |-\upalpha \rangle)/\sqrt{2}$, is prepared by sending a coherent state $\ket{\upalpha}$ through a Kerr cell, which is the nonclassical \cite{Rahimi-Keshari} and non-Gaussian \cite{Rahimi-Keshari-qpt} unitary operation $\widehat{U}_{\mathrm{cat}}=\mathrm{e}^{-i\pi(\widehat{\mathrm{a}}^{\dagger}\widehat{\mathrm{a}})^2/2}$. From  Ref.~\cite{Rahimi-Keshari}, we have
\begin{equation}
\mathpzc{E}_{\mathrm{cat}}[|\upalpha\rangle\langle\upalpha|] =\frac{1}{2}\big(\ket{\upalpha}+i\ket{-\upalpha}\big)\big(\bra{\upalpha}-i\bra{-\upalpha}\big).
\end{equation}
Hence, the normally-ordered moments for the output cat-state can be calculated from Eq.~\eqref{mom-out-coh} as
\begin{align}
M_{jk}(\mathpzc{E}_{\mathrm{cat}}[|\upalpha\rangle\langle\upalpha|])=&\frac{1}{2}\big( 1+(-1)^{j+k}+i[(-1)^j-(-1)^k]\big) \nonumber \\
& \times \upalpha^j\upalpha^{\ast k},
\end{align}
whence Eq.~\eqref{sup} yields
\begin{equation}
\mathpzc{E}_{jk}^{mn}=\frac{1}{2}\big( 1+(-1)^{j+k}+i[(-1)^j-(-1)^k]\big)\updelta_{mj}\updelta_{nk}.
\label{cat-gen-ten}
\end{equation}

%%%%%%%%%%%%%%%%%%%%%%%%%%%%%%%
\subsection{Noiseless linear amplifier}

Through an amplification process, one can enlarge the amplitude of the coherent state $\ket{\upalpha}$ to $\ket{g\upalpha}$, where $g>1$ is the amplification gain. It has been shown that \textit{deterministic} noiseless linear amplification (NLA) is not possible \cite{Caves}. However, NLA is feasible with some probability of success \cite{Ralph-Lund}, that is, $P_{\mathrm{succ}}=e^{-(1-g^2)|\upalpha|^2}/(g^2-1)^N$ assuming $N \gg g|\upalpha|$, where $N$ is the number of amplification units called \textit{quantum scissors} \cite{Ralph-Lund}. The action of the NLA on an input coherent state is as follows:
\begin{equation}
\mathpzc{E}_{\mathrm{NLA}}[\ketbra{\upalpha}{\upalpha}]= P_{\mathrm{succ}} \ketbra{g\upalpha}{g\upalpha} +(1- P_{\mathrm{succ}}) \ketbra{0}{0},
\end{equation}
where $\ket{0}$ is the vacuum state.  Equation~\eqref{mom-out-coh} then yields
\begin{equation}
M_{jk}(\mathpzc{E}_{\mathrm{NLA}}[\ketbra{\upalpha}{\upalpha}]) = g^{j+k}  P_{\mathrm{succ}}  \upalpha^j\upalpha^{\ast k},
\end{equation}
from which by using Eq.~\eqref{sup} we obtain
\begin{equation}
\mathpzc{E}_{jk}^{mn}=\frac{g^{j+k}(g^2-1)^{m-j-N}}{(m-j)!}\updelta_{m-j,n-k}.
\end{equation}
For an arbitrary input state $\widehat{\varrho}$, using Eq.~\eqref{mom-in-out}, the output  moments in terms of the input moments are determined
\begin{equation}
\label{M-out-NLA}
M_{jk}(\mathpzc{E}_{\mathrm{NLA}}[\widehat{\varrho}])= g^{j+k}  P_{\mathrm{succ}} M_{jk}(\widehat{\varrho}).
 \end{equation} 
 
Through this relation, one can check nonclassicality of the output states of the NLA. In order to examine quantum properties of a field, an appropriate measure is Mandel's $ Q$ parameter, which is defined as follows for the quantum state $\widehat{\varrho}$ \cite{Mandel}:
\begin{align}
Q:=  \frac{\langle \widehat{n}^2 \rangle - \langle \widehat{n}\rangle^2 }{\langle\widehat{n}\rangle}-1, 
\end{align}
where $\widehat{n}=\widehat{\mathrm{a}}^{\dagger}\widehat{\mathrm{a}}$ is the photon number operator and $\langle \circ\rangle :=\mathrm{Tr}[\widehat{\varrho}\circ]$> 
% [A]
%defines the expectation value over $\widehat{\varrho}$. 
For $ Q_{~\mathrm{NLA}}<0$ the output field is said to be sub-Poissonian, and therefore nonclassical \cite{Mandel}.  The Mandel parameter can also be written in the form of normally-order moments as 
\begin{align}
\label{Q-moment}
Q=   \frac{M_{22}(\widehat{\varrho}) - M_{11}^2(\widehat{\varrho})}{M_{11}(\widehat{\varrho})}. 
\end{align}
Thus, for the outputs of the NLA setup, by using Eqs.~\eqref{M-out-NLA} and \eqref{Q-moment}, we have
\begin{align}
 Q_{~\mathrm{NLA}}|_{\widehat{\varrho}}=\frac{g^2}{M_{11}(\widehat{\varrho})} \big[ M_{22}(\widehat{\varrho}) -   P_{\mathrm{succ}} M^2_{11}(\widehat{\varrho})\big].
\label{q-man-nnla}
\end{align}
As an example, let us check the nonclassicality for the coherent state $\ket{\upalpha}$ as the input of the NLA. Using Eq.~\eqref{q-man-nnla}, we  have
\begin{equation}
 Q_{~\mathrm{NLA}}|_{|\upalpha\rangle}=g^2 |\upalpha|^2 ( 1-  P_{\mathrm{succ}} ).
\end{equation}
This implies that, according to the $Q$ parameter, the output state is a classical state. This is, indeed, expected since the output is an amplified coherent state.

%%%%%%%%%%%%%%%%%%%%%%%%%%%%%%%%%%%%%
\squeezetable
\begin{table} 
\begin{ruledtabular}  
\caption{Superoperator elements for some quantum-optical processes.}        
\label{table-1}
\begin{tabular}{lcc}
\noalign{\vskip 0.13cm}
$\mathpzc{E}$ & $M_{jk}(\mathpzc{E}[\ket{\bm\upalpha}\bra{\bm\upalpha}])$ & $\mathpzc{E}_{\bm{jk}}^{\bm{mn}}$ \\[0.5ex]  
\hline    
\noalign{\vskip 0.1cm}
$\mathpzc{E}_{\mathrm{id}}$            & $\upalpha^j\upalpha^{\ast k}$    &   $\updelta_{mj}\updelta_{nk}$  \\ 
\\    
$\mathpzc{E}_{\mathrm{att}}$        & $\eta^{j+k}\upalpha^j\upalpha^{\ast k} $ & $\eta^{j+k}\updelta_{mj}\updelta_{nk}$  \\
\\
$\mathpzc{E}_{\mathrm{disp}}$    & $(\upalpha+\upbeta)^j(\upalpha^\ast+\upbeta^\ast)^k$   & $\frac{j!k!\upbeta^{j-m}\upbeta^{\ast k-n}}{m!n!(j-m)!(k-n)!}$  \\
\\
$\mathpzc{E}_{\mathrm{add}}$    & $\upalpha^{j+1}\upalpha^{\ast k+1}$ & $\updelta_{m,j+1}\updelta_{n,k+1}$  \\
                    {}                          & $+(k+j+1)\upalpha^{j}\upalpha^{\ast k}$ & $+(k+j+1)\updelta_{mj}\updelta_{nk}$ \\
                    {}                   &  $+kj\upalpha^{j-1}\upalpha^{\ast k-1}$ & $+kj\updelta_{m,j-1}\updelta_{n,k-1}$ \\
\\
$\mathpzc{E}_{\mathrm{sub}}$ & $\upalpha^{j+1}\upalpha^{\ast k+1}$ & $\updelta_{m,j+1}\updelta_{n,k+1}$ \\
\\
$\mathpzc{E}_{\mathrm{cat}}$  & $\frac{1}{2}\{1+(-1)^{j+k}+$ &   $\frac{1}{2}\{1+(-1)^{j+k}+$ \\
                          {}                    & $i[(-1)^j-(-1)^k]\}\upalpha^j \upalpha^{\ast k}$ & $i[(-1)^j-(-1)^k]\}\updelta_{mj}\updelta_{nk}$ \\
\\
$\mathpzc{E}_{\mathrm{dec}}$ & -- & $\frac{j!k!}{m!n!}\frac{N^{j-m}}{(j-m)!}[1-\nu^2(t)]^{j-m}$\\
                                   {}                                           &  {}  &       $\times \nu^{m+n}(t)\updelta_{j-m, k-n}     $ \\
\\
$\mathpzc{E}_{\mathrm{BS}}$ & $(T\upalpha_1-R\upalpha_2)^{j_1}(R\upalpha_1+T\upalpha_2)^{j_2}$ & $\sum_{pr}(-1)^{j_1+k_1-p-r} $ \\
              {}              & $\times (T\upalpha^\ast_1-R\upalpha^\ast_2)^{k_1}(R\upalpha^\ast_1+T\upalpha^\ast_2)^{k_2}$  &    $\times \binom{j_1}{p}\binom{k_1}{r}\binom{j_2}{m_1-p}\binom{k_2}{n_1-r}$ \\
							{}  &  {}  &   $\times T^{2p+2r+j_2+k_2-m_1-n_1}$ \\
							{}  &  {}  &   $\times R^{m_1+n_1+j_1+k_1-2p-2r}$ \\
							{}  &  {}  &   $\times \updelta_{m_1+m_2,j_1+j_2}$ \\
							{}  &  {}  &   $\times \updelta_{n_1+n_2,k_1+k_2}$
\\
$\mathpzc{E}_{\mathrm{NLA}}$ & $\frac{g^{j+k} \mathrm{e}^{-(1-g^2)|\upalpha|^2}}{(g^2-1)^N}\upalpha^j\upalpha^{\ast k}$ & $\frac{g^{j+k}(g^2-1)^{m-j-N}}{(m-j)! }$ \\
{} & {} &      $\times \updelta_{n-k, m-j}$   \\
\noalign{\vskip 0.1cm}
\end{tabular}
\end{ruledtabular}
\end{table}
%%%%%%%%%%%%%%%%%%%%%%%%%%%%%%%%%%%%%

%%%%%%%%%%%%%%%%%%%%%%%%%%%%%%%
\subsection{Decoherence}   
\label{decoherence}

Decoherence of an optical state, caused by a thermal bath with the mean photon number $N$, can be studied through the simple model of Ref. \cite{Richter}. According to this model, the characteristic function of the optical state $\widehat{\varrho}$ at time $\tau$ is given by
\begin{equation}
\Lambda_{\widehat{\varrho}}(\upxi, \tau)=\mathrm{e}^{-b(\tau)|\upxi|^2}\Lambda_{\widehat{\varrho}}(\upxi\nu(\tau)),
\label{charac-decoh}
\end{equation}
where $\Lambda_{\widehat{\varrho}}(\upxi)=\mathrm{Tr}[\widehat{\varrho}~ \mathrm{e}^{\upxi \widehat{\mathrm{a}}^{\dagger}}\mathrm{e}^{-\upxi^\ast\widehat{\mathrm{a}}}]$  is the characteristic function of the state $\widehat{\varrho}$, $b(\tau)=N[1-\nu^2(\tau)]$, and $\nu(\tau)=\mathrm{e}^{-\upgamma \tau}$ with $\upgamma$ being the damping rate \cite{Cahill-Glauber}. Employing Eqs.~(\ref{sup-alternate}) and (\ref{charac-decoh}), the time-dependent superoperator tensor of this process is obtained as
\begin{align}
\label{sup-decoh}
\mathpzc{E}_{jk}^{mn}(\tau)=&\frac{j!k!}{m!n!} \frac{ N^{j-m}}{(j-m)!} \nu^{m+n}(\tau)[1-\nu^2(\tau)]^{j-m} \nonumber \\
&\times \updelta_{j-m, k-n}. 
\end{align}
Inserting Eq.~\eqref{sup-decoh} into Eq.~\eqref{mom-in-out}, we obtain the elements of the normally-ordered moments at time $\tau$,  
\begin{align}
\label{nom-at-t}
M_{jk}(\tau)=&j!k!\sum_{m=0}\frac{ N^{j-m}}{(j-m)!}\nu^{2m-j+k}(\tau)\nonumber \\
&\times [1-\nu^2(\tau)]^{j-m}M_{m,m-j+k}(0),
\end{align}
where $M_{jk}(0)$ is the $jk$th moment of the state at time $\tau=0$. 

Of particular interest is the decoherence of a squeezed vacuum state in a thermal bath. In order to study deformation of this state, a simple way is to check the variance of one of the field quadratures, say $\widehat{x}$, as a function of time,
\begin{equation}
\Delta_x(\tau)= \langle \widehat{x}^2(\tau) \rangle-\langle \widehat{x}(\tau) \rangle^2.
\end{equation}
The above quantity can also be expressed in terms of the normally-ordered moments as follows:
\begin{equation} 
\Delta_x(\tau)= \frac{1}{4}\big[1+2M_{11}(\tau)+M_{20}(\tau)+M^{\ast}_{20}(\tau)\big],
\end{equation}
where we have used the relation between the field quadratures and annihilation/creation operators---$\widehat{x}=(\widehat{\mathrm{a}}+\widehat{\mathrm{a}}^{\dagger})/2$. 

Now by using Eq.~\eqref{nom-at-t}, we obtain the variance of the quadrature $\widehat{x}$ at time $\tau$ as
\begin{equation}
\Delta_x(\tau)=\frac{1}{4}+\frac{N}{2} (1-\mathrm{e}^{-2\upgamma \tau})  .
\end{equation}
It is evident that in the limit of $\tau\rightarrow 0$ the variance converges to $1/4+N/2$, which depends explicitly on $N$. This implies that at sufficiently large times the squeezed vacuum state becomes a thermal state with  $N/2$ of added noise.  Stronger thermal baths, i.e., larger amounts of $N$, results in adding more noise to the state.  

\section{Gaussian Processes}
\label{gaussian Processes}

In quantum optics, a Gaussian process/channel is defined as a process that maps any Gaussian state to another Gaussian state \cite{Ferraro,Adesso}. This class of channels includes attenuation, amplification, thermalization, displacement, and squeezing, to name a few. Such channels are of wide interest, e.g., in implementation of continuous-variable quantum information protocols including quantum key distribution and teleportation \cite{Adesso,Weedbrook}. 

Similarly multi-mode Gaussian processes (MMGPs), such as beam splitters and multimode squeezers, map $\mathpzc{m}$-mode Gaussian state to $\mathpzc{m}$-mode Gaussian state. An $\mathpzc{m}$-mode optical state ${\widehat{\varrho}_G}$ is called Gaussian if its Wigner function on the quantum phase space has a Gaussian form \cite{Ferraro},
\begin{equation}
W_G(R)= \frac{1}{ \sqrt{(2 \pi)^{2\mathpzc{m}}\det[V ]}} \mathrm{e}^{{-\frac{1}{2}(R- \overline{R})^{T}V^{-1} (R-\overline{R})}},
\label{Wig-Gauss}
\end{equation}
where $R=(x_1~p_1~\ldots~x_{\mathpzc{m}}~p_{\mathpzc{m}})^T$ and $\overline{R}=(\overline{x}_1~\overline{p}_1~\ldots~\overline{x}_{\mathpzc{m}}~\overline{p}_{\mathpzc{m}})^T$ are, respectively, the vector of the quadratures and the average value of the quadratures. The associated quadrature operators 
% [A]
%of $x_j$ and $p_j$ 
are defined as $\widehat{x}_j=(\widehat{\mathrm{a}}_j+ \widehat{\mathrm{a}}_j^{\dag})/2$ and $\widehat{p}_j=(\widehat{\mathrm{a}}_j-\widehat{\mathrm{a}}_j^{\dag})/(2i)$. Here $V $ is a $2\mathpzc{m}\times 2\mathpzc{m}$ real symmetric matrix, called the covariance matrix,
\begin{equation}
\label{CM-elements}
V_{jk}= \frac{1}{2}\langle \widehat{R}_j\widehat{R}_k+ \widehat{R}_k \widehat{R}_j\rangle_G -\langle \widehat{R}_j\rangle_G \langle \widehat{R}_k\rangle_G,
\end{equation}
where $\langle \circ \rangle_G:=\mathrm{Tr}[\widehat{\varrho}_G ~\circ]$.

In order to characterize an MMGP $\mathpzc{E}$ by the method described in Sec. \ref{Formalism}, one first needs to send multi-mode coherent states $|\bm{\upalpha}\rangle=|\upalpha_1,\upalpha_2,\ldots,\upalpha_{\mathpzc{m}}\rangle$ as inputs to the process. Next one should measure the output moments $M_{\mathbf{j}\mathbf{k}} (\mathpzc{E}[\ketbra{\bm{\upalpha}}{\bm{\upalpha}}])$. We remind that $\mathpzc{m}$-mode normally-ordered moments for a quantum state $\widehat{\varrho}$ are given by (cf. Eq.~(\ref{1-moments}))
\begin{align}
M_{j_1k_1,\dots, j_{\mathpzc{m}}k_{\mathpzc{m}}}(\widehat{\varrho}) = \mathrm{Tr}[\widehat{\varrho}~\widehat{\mathrm{a}}^{\dag k_1}_1 \widehat{\mathrm{a}}^{j_1}_1 \dots 
\widehat{\mathrm{a}}^{\dag k_{\mathpzc{m}}}_{\mathpzc{m}} \widehat{\mathrm{a}}^{j_{\mathpzc{m}}}_{\mathpzc{m}} ].
\end{align} 
After obtaining the moments, one should evaluate the superoperator elements through Eq.~\eqref{mult-sup}. 

Our objective here is to show how we can obtain the rank-$4$ tensor of a Gaussian process, in addition, we estimate the required resource (the number of input multi-mode coherent state) for this task. To this end, we first need to remind two pertinent points. (i) For any Gaussian state, all normally-ordered moments can be written in terms of the mean-value vector and the covariance matrix of the state; that is only two moments would suffice. For example, for a two-mode Gaussian state one can restate the moment $M_{10,01}\equiv  \langle\widehat{\mathrm{a}}\, \widehat{\mathrm{b}}^{\dagger}\rangle_G$ as $M_{10,01}=(V_{13}+V_{24}- iV_{14}+iV_{23})+ (\overline{x}_1 \overline{x}_2 + \overline{p}_1 \overline{p}_2 - i \overline{x}_1 \overline{p}_2  +i \overline{p}_1 \overline{x}_2)$. (ii) The action of any trace-preserving MMGP can be described by the following transformations \cite{Weedbrook}:
\begin{align}
\overline{R}(\mathpzc{E}[\widehat{\varrho}_G]) &=S\overline{R}(\widehat{\varrho}_G)+D, \label{MMGP-mean}\\
V(\mathpzc{E}[\widehat{\varrho}_G]) &= SV(\widehat{\varrho}_G) S^{T}+ E,
\label{MMGP-CM}
\end{align}
where $S$ and $E=E^T$ are $2\mathpzc{m}\times 2\mathpzc{m}$ real matrices, and $D\in \mathbb{R}^ {2\mathpzc{m}}$. The triplet $(S,E,D)$ completely specifies an MMGP, which implies that in  general the number of unknown parameters of a Gaussian process is $3\mathpzc{m} \times (2\mathpzc{m}+1)$. 

By probing each multi-mode coherent state, Eq.~\eqref{MMGP-mean} yields $2\mathpzc{m}$ equations; thus, probing $2\mathpzc{m}+1$ different multi-mode coherent states provides $2\mathpzc{m} \times (2\mathpzc{m}+1)$ equations, which is equal to the number of unknown parameters of $S$ and $D$. One then can find $E$ from Eq.~\eqref{MMGP-CM}. More specifically, by employing Eq.~\eqref{MMGP-mean} one can find $S$ and $D$ through the following equation:
\begin{equation}
\overset{\rightharpoonup}{\mathpzc{S}_j}=\overline{\mathpzc{R}}^{-1} \overline{\mathpzc{R}}_{j}^{\mathrm{out}},~~~~~~~~~j=1,\ldots,2\mathpzc{m}+1,
\end{equation}
where $\overset{\rightharpoonup}{\mathpzc{S}_j}=(S_j~D_j)^T$ is built by merging the $j$th row of the matrix $S$, $S_j$, and the $j$th element of the vector $D$, and
\begin{equation}
\overline{\mathpzc{R}}=
\left(\begin{array}{ccccc}
\overline{R}_{1}^{(1)} & \overline{R}_{2}^{(1)} & \ldots& \overline{R}_{2\mathpzc{m}}^{(1)}  & 1   \\
\overline{R}_{1}^{(2)} & \overline{R}_{2}^{(2)} & \ldots & \overline{R}_{2\mathpzc{m}}^{(2)} &  1   \\ 
\vdots  \\
\overline{R}_{1}^{(2\mathpzc{m}+1)} & \overline{R}_{2}^{(2\mathpzc{m}+1)} & \ldots & \overline{R}_{2\mathpzc{m}}^{(2\mathpzc{m}+1)} & 1
\end{array}\right) \nonumber
\end{equation}
is the \textit{matrix of resources}. Note that $\overline{\mathpzc{R}}$ is a real matrix that includes the mean value of the quadratures of different input coherent states, in which $\overline{R}_{i}^{(k)}$ is the $i$th quadrature of the $k$th input coherent state.
Besides, $\overline{\mathpzc{R}}_{j}^{\mathrm{out}}= (\overline{R}_{j}^{\mathrm{out}~(1)}~~ \overline{R}_{j}^{\mathrm{out}~ (2)}~~\ldots~~\overline{R}_{j}^{\mathrm{out}~(2\mathpzc{m}+1)})^T$ represents the mean values of the quadratures of the output states, which are to be estimated from experiment. Evidently, for the purpose of determining the unknowns, the matrix $\overline{\mathpzc{R}}$ needs to be invertible, that is, the input coherent states should be chosen such that they meet this condition. 

% [A] removed: seemed redundant and repetitive
%\begin{proposition}
%To characterize an $\mathpzc{m}$-mode Gaussian process described by Eqs.~(\ref{MMGP-mean}) and (\ref{MMGP-CM}), $2\mathpzc{m}+1$ different input $\mathpzc{m}$-mode coherent states are needed.
%\end{proposition}

This result that $2\mathpzc{m}+1$ different input $\mathpzc{m}$-mode coherent states suffice offers a relative improvement on the existing methods for csQPT \citep{Lobino,Rahimi-Keshari-qpt}, where precise identification of a process has been argued to require infinite number of coherent states. Note that our result agrees with the proposed QPT method using the Husimi function \cite{Wang-csQPT}.

As we argued, Gaussian process tomography is achieved by evaluating the triplet $(S,N,D)$. This can also be seen more explicitly through expressing the tensor of a Gaussian process in terms of the triplet elements, 
% [A]
%. To do so, let us focus on the simple 
e.g., in the 
case of the single-mode Gaussian process $\mathpzc{E}$, for which from Eq.~\eqref{MMGP-mean} and by assuming a coherent inout state $|\upalpha\rangle$ we have
% [A] removed: for brevity
%\begin{align}
%\overline{x}_1(\mathpzc{E}[\widehat{\varrho}])= &S_{11} \overline{x}_1(\widehat{\varrho}) + S_{12} \overline{p}_1(\widehat{\varrho}) +D_1, \nonumber \\
%\overline{p}_1(\mathpzc{E}[\widehat{\varrho}])= &S_{21} \overline{x}_1(\widehat{\varrho}) + S_{22} \overline{p}_1(\widehat{\varrho}) +D_2.
%\end{align}
%Since we assume our input to be the coherent state $\widehat{\varrho}=\ketbra{\upalpha}{\upalpha}$, the above relations become 
\begin{align}
\overline{x}_1(\mathpzc{E}[\ketbra{\upalpha}{\upalpha}])= & \frac{S_{11}}{2}(\upalpha +\upalpha^\ast) + \frac{S_{12}}{2i}(\upalpha -\upalpha^\ast) +D_1,  \\
\overline{p}_1(\mathpzc{E}[\ketbra{\upalpha}{\upalpha}])= & \frac{S_{21}}{2}(\upalpha +\upalpha^\ast) + \frac{S_{22}}{2i}(\upalpha -\upalpha^\ast) +D_2.
\end{align}
Hence
\begin{align}
M_{10}(\mathpzc{E}[\ketbra{\upalpha}{\upalpha}]) 
% [A] removed: redundant
%= &  \overline{x}_1(\mathpzc{E}[\ketbra{\upalpha}{\upalpha}]) +i \overline{p}_1(\mathpzc{E}[\ketbra{\upalpha}{\upalpha}])  \nonumber \\
= & (1/2)(S_{11}+iS_{21}) (\upalpha +\upalpha^\ast)  \nonumber \\
& + (1/2)(S_{22}-iS_{12}) (\upalpha -\upalpha^\ast) ~ \nonumber \\
& +(D_1+iD_2),  \\
% [A] removed: redundant
%M_{01}(\mathpzc{E}[\ketbra{\upalpha}{\upalpha}]) = &   M_{10}^{\ast}(\mathpzc{E}[\ketbra{\upalpha}{\upalpha}])  ,
%~ \nonumber \\
M_{11}(\mathpzc{E}[\ketbra{\upalpha}{\upalpha}])= & \big[V_{11}(\mathpzc{E}[\ketbra{\upalpha}{\upalpha}])+ V_{22}(\mathpzc{E}[\ketbra{\upalpha}{\upalpha}]) \nonumber \\
& -1\big] + |M_{10}(\mathpzc{E}[\ketbra{\upalpha}{\upalpha}]) |^2,  \\
M_{20}(\mathpzc{E}[\ketbra{\upalpha}{\upalpha}])= & \big[V_{11}(\mathpzc{E}[\ketbra{\upalpha}{\upalpha}]) - V_{22}(\mathpzc{E}[\ketbra{\upalpha}{\upalpha}]) \nonumber \\
& +2i V_{12}(\mathpzc{E}[\ketbra{\upalpha}{\upalpha}])\big] \nonumber \\
& + M_{10}^2 (\mathpzc{E}[\ketbra{\upalpha}{\upalpha}]), %\\
% [A] removed: redundant
%M_{02}(\mathpzc{E}[\ketbra{\upalpha}{\upalpha}]) =& M_{20}^{\ast}(\mathpzc{E}[\ketbra{\upalpha}{\upalpha}])  ,
\end{align}
where $V(\mathpzc{E}[\ketbra{\upalpha}{\upalpha}])$ is the covariance matrix of the output state. 
% [A]
% corresponding to the input coherent state $|\upalpha\rangle$. 
Note that using Eq.~\eqref{MMGP-CM}, we can conclude that $V_{jk}(\mathpzc{E}[\ketbra{\upalpha}{\upalpha}])$s are $\alpha$-independent, because $V(\ketbra{\upalpha}{\upalpha}])= \openone/4$, where $\openone$ is the identity operator. Now the tensor elements $\mathpzc{E}^{mn}_{jk}$ can be calculated from Eq.~\eqref{sup}; e.g., 
\begin{align}
\mathpzc{E}^{00}_{10} &=  D_1 + i D_2,\\
\mathpzc{E}^{10}_{10} &=  \frac{1}{2} [S_{11} + -i (S_{12}- S_{21}) + S_{22}].
% [A] there seems a typo in the above equation: + - 
\end{align}
This completes obtaining the superoperator of the single-mode Gaussian process. Extension to multi-mode Gaussian processes is straightforward noting Eqs.~\eqref{MMGP-mean} and \eqref{mult-sup}. 
% [A] redundant and repetitive
%Thus as a result, by using only $2\mathpzc{m}+1$ input coherent states, one can perform a complete tomography of a Gaussian quantum process. 

As a final remark, we stress that in this section it has been assumed that the quantum process under study is known to be Gaussian. This prior partial information simplified the analysis of the process characterization by restricting only to two moments. When this information is not given or when there is some noise along with a Gaussian process, the simplified method of this section would not work and we should resort to the general method of Sec.~\ref{Formalism}. In fact, in non-Gaussian or noisy Gaussian processes, if we measure only the first and second moments, we would lose a part of the information about the process and thus the characterization will bear errors. It is an interesting and relevant question---but beyond the goal of this paper---to compute this error and analyze its behavior.

%%%%%%%%%%%%%%%%%%%%%%%%%%%%%%%%%%%%%
\section{Summary}
\label{Conclusion}

We have laid out a method for coherent-state quantum process tomography based on measurement of normally-ordered moments in order to characterize an unknown quantum(-optical) process. This method may be suitable in particular for characterization of processes for which it is somehow known that a finite number of moments suffices, whereas even in such cases most existing methods may still require relatively more resources or measurements for full characterization. We have demonstrated utility of our method through complete characterization of several quantum-optical processes and Gaussian processes.  

It should be noted that for characterization of general processes or even Gaussian processes accompanied with noise or imperfections, our method may not offer advantages because in such cases one may need to measure a large number of moments to obtain a reasonable characterization of the process with limited error. Given this point, it still remains (for a further study) to do a comprehensive and comparative analysis of errors and imperfection vs. resources for coherent-state process tomography schemes including ours.

\begin{acknowledgments}
M.G. acknowledges initial inputs of S. Rahimi-Keshari. Partial support by Sharif University of Technology's Office of Vice President for Research (to A.T.R.) is also acknowledged.
\end{acknowledgments}

%%%%%%%%%%%%%%%%%%%%%%%%%%%%%%%%%%%%%

%%%%%%%%%%%%%%%%%%%%%%%%%%%%%%%%%%%%%

%\twocolumngrid
\newpage
\begin{widetext}

\appendix 

\section{Proof that the power series of $M_{jk}(\mathpzc{E}[\ketbra{\upalpha}{\upalpha}])$ is convergent} 
\label{conv-of-M}

Using the expansion of a coherent state $\ket{\upalpha}$ in the Fock basis and according to the definition of the output moments in Eq.~\eqref{mom-out-coh}, we have
\begin{align}
M_{jk}(\mathpzc{E}[\ketbra{\upalpha}{\upalpha}]) =&  \sum_{m,n=0}^{\infty} \varrho_{mn}(\upalpha) \text{Tr}\left[ \mathpzc{E}[\ketbra{m}{n}] \widehat{\mathrm{a}}^{\dag k} \widehat{\mathrm{a}}^j\right]   \nonumber\\
=& \sum_{m,n=0}^{\infty} \varrho_{mn}(\upalpha)   \sum_{\ell=0}^{\infty} \bra{\ell} \widehat{\mathrm{a}}^j\mathpzc{E}[\ketbra{m}{n}] \widehat{\mathrm{a}}^{\dag k}\ket{\ell} \nonumber\\
=& \sum_{m,n=0}^{\infty} \varrho_{mn}(\upalpha) \sum_{\ell=0}^{\infty}\sqrt{\frac{(\ell+j)!(\ell+k)!}{(\ell!)^2}}  \bra{\ell+j}\mathpzc{E}[|m\rangle\langle n|] |\ell+k\rangle,
\end{align}
where $\varrho_{mn}(\upalpha)= \mathrm{e}^{-|\upalpha|^2}{\upalpha^m {\upalpha^\ast}^n}/{\sqrt{m! n!}}$ and in the last part we have used the relation $\widehat{\mathrm{a}}^{\dag k}\ket{\ell}=\sqrt{(\ell+k)!/\ell!}\ket{\ell+k}$. The output moments are given by the elements of the superoperator in the Fock basis $\mathcal{E}_{jk}^{^{mn}}:=\bra{j}\mathpzc{E}[|m\rangle\langle n|] |k\rangle$ \cite{Rahimi-Keshari-qpt}, as  
\begin{align}
\label{mom-in-Fock}
M_{jk}(\mathpzc{E}[\ketbra{\upalpha}{\upalpha}]) =&  \sum_{m,n=0}^{\infty} \varrho_{mn}(\upalpha)  \sum_{\ell=0}^{\infty}\sqrt{\frac{(j+\ell)!(k+\ell)!}{(\ell!)^2}}  \mathcal{E}_{j+\ell,k+\ell}^{^{mn}},
\end{align}
which clearly is an entire function. Thus, $M_{jk}(\mathpzc{E}[\ketbra{\upalpha}{\upalpha}])$ can be expressed as a power series that is convergent. Note that the second part does not depend on $\upalpha$. 
In addition, using Eq.~\eqref{mom-in-Fock}, we can find a relation between the superoperator elements in the moment basis and in the Fock basis as
\begin{equation}
\label{tensor-both}
\mathpzc{E}_{jk}^{mn}=\sum_{\ell,s=0}^{\infty}\frac{(-1)^s}{s!\ell!}\sqrt{\frac{(j+\ell)!(k+\ell)!}{(m-s)!(n-s)!}} \mathcal{E}_{j+\ell,k+\ell}^{m-s,n-s}.
\end{equation}
Since the coefficients of $\mathcal{E}$ on the right-hand side are nonzero, one can invert Eq.~\eqref{tensor-both} and find the tensor of the process in the Fock basis in terms of the tensor elements in the moment basis.

%%%%%%%%%%%%%%%%%%%%%%%%%%%%%%%%%%%%%%
\section{Process tensor for a beam splitter} 
\label{app-BS}

In order to obtain the tensor elements of a beam splitter, we first calculate the normally-ordered moments of the output state 
\begin{eqnarray}
M_{j_1j_2k_1k_2}(\mathpzc{E}_{\mathrm{BS}}[\ket{\upalpha_1, \upalpha_2}\bra{\upalpha_1, \upalpha_2}]) &=&\mathrm{Tr} \big[\mathpzc{E}_{\mathrm{BS}}[\ket{\upalpha_1, \upalpha_2}\bra{\upalpha_1, \upalpha_2}] \widehat{\mathrm{a}}_2^{\dag k_2}\widehat{\mathrm{a}}_1^{\dag k_1}\widehat{\mathrm{a}}_2^{j_2}\widehat{\mathrm{a}}_1^{j_1}\big]
\nonumber\\
&=&\mathrm{Tr}\big[ \ket{T \upalpha_1-R  \upalpha_2, R  \upalpha_1+T \upalpha_2}\bra{T \upalpha_1-R  \upalpha_2, R  \upalpha_1+T \upalpha_2} \widehat{\mathrm{a}}_2^{\dag k_2}\widehat{\mathrm{a}}_1^{\dag k_1}\widehat{\mathrm{a}}_2^{j_2}\widehat{\mathrm{a}}_1^{j_1}\big]
\nonumber\\
&=& (T \upalpha_1-R  \upalpha_2)^{j_1}(R  \upalpha_1+T \upalpha_2)^{j_2}(T {\upalpha^\ast}_1-R  {\upalpha^\ast}_2)^{k_1} (R  {\upalpha^\ast}_1+ T  {\upalpha^\ast}_2)^{k_2},
\end{eqnarray}
then, by applying Eq.~\eqref{mult-sup}, we obtain the following expression 
\begin{equation}
\mathpzc{E}_{j_1j_2k_1k_2}^{m_1m_2n_1n_2}=\frac{1}{m_1! m_2! n_1! n_2!}\partial_{\upalpha_1}^{m_1} \partial_{\upalpha^\ast_1}^{n_1} \partial_{\upalpha_2}^{m_2} \partial_{\upalpha^\ast_2}^{n_2} \left[  (T \upalpha_1-R  \upalpha_2)^{j_1}(R  \upalpha_1+T \upalpha_2)^{j_2}(T {\upalpha^\ast}_1-R  {\upalpha^\ast}_2)^{k_1} (R  {\upalpha^\ast}_1+ T  {\upalpha^\ast}_2)^{k_2} \right]\Big|_{\upalpha_1, \upalpha_2=0}.
\end{equation}
Next, by using the binomial expansion, the tensor elements are obtained as 
\begin{eqnarray}
\mathpzc{E}_{j_1j_2k_1k_2}^{m_1m_2n_1n_2} &=&\frac{1}{m_1!m_2!n_1!n_2!}\sum_{p,q,r,s}\binom{j_1}{p}\binom{k_1}{r}\binom{j_2}{q}\binom{k_2}{s} (-1)^{j_1+k_1-p-r} T ^{j_2+k_2+p+r-q-s}\nonumber\\
&&\times  R  ^{j_1+k_1+q+s-p-r} \partial_{\upalpha_1}^{m_1} \partial_{\upalpha^\ast_1}^{n_1} \partial_{\upalpha_2}^{m_2} \partial_{\upalpha^\ast_2}^{n_2} 
\left[ \upalpha_1^{p+q} \upalpha_2^{j_1+j_2-p-q} {\upalpha^\ast}_1^{r+s} {\upalpha^\ast}_2^{k_1+k_2-r-s}\right]\Big|_{\upalpha_1, \upalpha_2=0}
\nonumber\\
&=&\sum_{p,r}\binom{j_1}{p}\binom{k_1}{r}\binom{j_2}{m_1-p}\binom{k_2}{n_1-r}  (-1)^{j_1+k_1-p-r} T ^{2p+2r+j_2+k_2-m_1-n_1}
\nonumber\\
&&\times R  ^{m_1+n_1+j_1+k_1-2p-2r}\delta_{m_1+m_2,j_1+j_2}\delta_{n_1+n_2,k_1+k_2}.
\end{eqnarray}

\twocolumngrid
\end{widetext}


\begin{thebibliography}{99}

\bibitem{Nielsen} M. A. Nielsen and I. L. Chuang, \textit{Quantum Computation and Quantum Information} (Cambridge University Press, Cambridge, 2000). 

\bibitem{Chuang}
I. L. Chuang and M. A. Nielsen, J. Mod. Opt. \textbf{44}, 2455 (1997). 
% Prescription for experimental determination of the dynamics of a quantum black box

\bibitem{Poyatos}
J. F. Poyatos, J. I. Cirac, and P. Zoller, Phys. Rev. Lett. \textbf{78}, 390 (1997).
% Complete Characterization of a Quantum Process: The Two-Bit Quantum Gate

\bibitem{Leung}
D. W. Leung, J. Math. Phys. \textbf{44}, 528 (2003).
% Choi's proof as a recipe for quantum process tomography

\bibitem{D'Ariano-AAPT-1}
G. M. D'Ariano and P. Lo Presti, Phys. Rev. Lett. \textbf{86}, 4195 (2001).
% Quantum Tomography for Measuring Experimentally the Matrix Elements of an Arbitrary Quantum Operation

\bibitem{Altepeter}
J. B. Altepeter, D. Branning, E. Jeffrey, T. C. Wei, P. G. Kwiat, R. T. Thew, J. L. O'Brien, M. A. Nielsen, and A. G. White, Phys. Rev. Lett. \textbf{90}, 193601 (2003).
% Ancilla-Assisted Quantum Process Tomography

\bibitem{D'Ariano-AAPT-2}
G. M. D'Ariano and P. Lo Presti, Phys. Rev. Lett. \textbf{91}, 047902 (2003).
% Imprinting Complete Information about a Quantum Channel on its Output State

\bibitem{Mohseni-DCQD}
M. Mohseni and D. A. Lidar, Phys. Rev. Lett. \textbf{97}, 170501 (2006).
% Direct Characterization of Quantum Dynamics

\bibitem{Mohseni}
M. Mohseni, A. T. Rezakhani, and D. A. Lidar, Phys. Rev. A \textbf{77}, 032322 (2008).
% Quantum-process tomography: Resource analysis of different strategies

\bibitem{Knill}
E. Knill, R. Laflamme, and G. J. Milburn, Nature \textbf{409}, 46 (2000).
% A scheme for efficient quantum computation with linear optics

\bibitem{Gisin}
N. Gisin, G. Ribordy, W. Tittel, and H. Zbinden, Rev. Mod. Phys. \textbf{74}, 145 (2002).
% Quantum cryptography

\bibitem{O'Brien}
J. L. O'Brien, G. J. Pryde, A. Gilchrist, D. F. V. James, N. K. Langford, T. C. Ralph, and A. G. White, Phys. Rev. Lett. \textbf{93}, 080502 (2004).
% Quantum Process Tomography of a Controlled-NOT Gate

\bibitem{Childs}
A. M. Childs, I. L. Chuang, and D. W. Leung, Phys. Rev. A \textbf{64}, 012314 (2001).
% Realization of quantum process tomography in NMR

\bibitem{Mitchell}
M. W. Mitchell, C. W. Ellenor, S. Schneider, and A. M. Steinberg, Phys. Rev. Lett. \textbf{91}, 120402 (2003).
% Diagnosis, Prescription, and Prognosis of a Bell-State Filter by Quantum Process Tomography

\bibitem{Lobino}
M. Lobino, D. Korystov, C. Kupchak, E. Figueroa, B. C. Sanders, and A. I. Lvovsky, Science \textbf{322}, 563 (2008).
% Complete Characterization of Quantum-Optical Processes

\bibitem{Rahimi-Keshari-qpt}
S. Rahimi-Keshari, A. Scherer, A. Mann, A. T. Rezakhani, A. I. Lvovsky, and B. C. Sanders, New J. Phys. \textbf{13}, 013006 (2011).
% Quantum process tomography with coherent states

\bibitem{Blandino}
R. Blandino, F. Ferreyrol, M. Barbieri, P. Grangier, and R. Tualle-Brouri, arXiv:1105.5510.
% A method for characterizing coherent-state quantum gates

\bibitem{Anis}
A. Anis and A. I. Lvovsky, New J. Phys. \textbf{14}, 105021 (2012).
% Maximum-likelihood coherent-state quantum process tomography 

\bibitem{Kumar}
R. Kumar, E. Barrios, C. Kupchak, and A. I. Lvovsky, Phys. Rev. Lett. \textbf{110}, 130403 (2013).
% Experimental Characterization of Bosonic Creation and Annihilation Operators

\bibitem{Wang-csQPT}
X. B. Wang, Z. W. Yu, J. Z. Hu, A. Miranowicz, and F. Nori, Phys. Rev. A \textbf{88}, 022101 (2013).
% Efficient tomography of quantum-optical gaussian processes probed with a few coherent states

\bibitem{Leonhardt}
U. Leonhardt, \textit{Measuring the Quantum State of Light} (Cambridge University Press, Cambridge, 1997).

\bibitem{Lvovsky}
A. I. Lvovsky and M. G. Raymer, Rev. Mod. Phys. \textbf{81}, 299 (2009).
% Continuous-variable optical quantum-state tomography

\bibitem{Vogel-Book}
W. Vogel and D. G. Welsch, \textit{Quantum Optics} (Wiley-VCH, Weinheim, 2006). 
% Quantum Optics

\bibitem{Rahimi-Keshari}
S. Rahimi-Keshari, T. Kiesel, W. Vogel, S. Grandi, A. Zavatta, and M. Bellini, Phys. Rev. Lett. \textbf{110}, 160401 (2013).
% Quantum Process Nonclassicality

\bibitem{Shchukin-1}
E. Shchukin, T. Richter, and W. Vogel, Phys. Rev. A \textbf{71}, 011802(R) (2005).
% Nonclassicality criteria in terms of moments

\bibitem{Shchukin-2}
E. V. Shchukin and W. Vogel, Phys. Rev. A \textbf{72}, 043808 (2005).
% Nonclassical moments and their measurement

\bibitem{Ferraro}
A. Ferraro, S. Olivares, and M. G. A. Paris, arXiv:quant-ph/0503237.
% gaussian states in continuous variable quantum information

\bibitem{Adesso}
G. Adesso and F. Illuminati, J. Phys. A: Math. Theor. \textbf{40}, 7821 (2007). 
% Entanglement in continuous-variable systems: recent advances and current perspectives

\bibitem{Weedbrook}
C. Weedbrook, S. Pirandola, R. Garcia-Patron, N. J. Cerf, T. C. Ralph, J. H. Shapiro, and S. Lloyd, Rev. Mod. Phys. \textbf{84}, 621 (2012).
% gaussian quantum information

\bibitem{Opatrny}
T. Opatrny, N. Korolkova, and G. Leuchs, Phys. Rev. A \textbf{66}, 053813 (2002).
% Mode structure and photon number correlations in squeezed quantum pulses

\bibitem{Pinel}
O. Pinel, P. Jian, R. M. de Araujo, J. Feng,  B. Chalopin, C. Fabre, and N. Treps, Phys. Rev. Lett. \textbf{108}, 083601 (2012).
% Generation and Characterization of Multimode Quantum Frequency Combs

\bibitem{Glauber} 
R. J. Glauber, Phys. Rev. Lett. \textbf{10}, 84 (1963).
% The Quantum Theory of Optical Coherence

\bibitem{Sudarshan}
E. C. G. Sudarshan, Phys. Rev. Lett. \textbf{10}, 277 (1963).
% Equivalence of Semiclassical and Quantum Mechanical Descriptions of Statistical Light Beams

\bibitem{Cahill-Glauber}
K. E. Cahill and R. J. Glauber, Phys. Rev. \textbf{177}, 1882 (1969).
% Density Operators and Quasiprobability Distributions

\bibitem{Scully}
M. O. Scully and M. S. Zubairy, \textit{Quantum Optics} (Cambridge University Press, Cambridge, 2001). 

\bibitem{Caves}
C. M. Caves, Phys. Rev. D \textbf{23}, 1693 (1981).
% Quantum-mechanical noise in an interferometer

\bibitem{Ralph-Lund}
T. C. Ralph and A. P. Lund, arXiv:0809.0326
% Nondeterministic Noiseless Linear Amplification of Quantum Systems

\bibitem{Mandel}
L. Mandel, Opt. Lett. \textbf{4}, 205 (1979).
% Sub-Poissonian photon statistics in resonance fluorescence

\bibitem{Richter}
T. Richter and W. Vogel, Phys. Rev. A \textbf{76}, 053835 (2007).
% Nonclassical characteristic functions for highly sensitive measurements

\end{thebibliography}
\end{document}